\documentclass[twocolumn,showpacs,preprintnumbers,prl,amsmath,amssymb]{revtex4}

\usepackage{graphicx}
\usepackage{bm}

\newcommand{\Srn}{SrFe$_{2-x}$Ni$_x$As$_2$}

\newcommand{\Ban}{BaFe$_{2-x}$Ni$_x$As$_2$}
\newcommand{\Bac}{BaFe$_{2-x}$Co$_x$As$_2$}
\newcommand{\Barh}{BaFe$_{2-x}$Rh$_x$As$_2$}
\newcommand{\Bapd}{BaFe$_{2-x}$Pd$_x$As$_2$}
\newcommand{\Sr}{SrFe$_2$As$_2$}
\newcommand{\Ba}{BaFe$_2$As$_2$}

\newcommand{\Bapt}{BaFe$_{1.90}$Pt$_{0.10}$As$_2$}
\newcommand{\Bap}{BaFe$_{2-x}$Pt$_{x}$As$_2$}

\newcommand{\tc}{$T_c$}

\begin{document}

\date{\today}

\title{Superconductivity at 23 K in Pt doped BaFe$_2$As$_2$ single crystals}

\author{S.~R.~Saha, T.~ Drye,  K.~Kirshenbaum, N.~P.~Butch,}
\author{Johnpierre~Paglione}

\email{paglione@umd.edu}

\address{Center for Nanophysics and
Advanced Materials, Department of Physics, University of Maryland,
College Park, MD 20742, USA.}

\author{P. Y. Zavalij}
\address{Department of Chemistry and Biochemistry, University of
Maryland, College Park, MD 20742, USA.}

\begin{abstract}
We report superconductivity in single crystals of the new
iron-pnictide system \Bapt\ grown by a self-flux solution method and
characterized via x-ray, transport, magnetic and thermodynamic
measurements. The magnetic ordering associated with a structural
transition at $139$~K present in \Ba\ is completely suppressed by
substitution of 5\% Fe with Pt and superconductivity is induced at a
critical temperature \tc = 23~K. Full diamagnetic screening in the
magnetic susceptibility and a jump in the specific heat at \tc\
confirm the bulk nature of the superconducting phase. All properties
of the superconducting state -- including transition temperature
\tc, the lower critical field $H_{c1}=200$~mT, upper critical field
$H_{c2} \approx$ 65~T, and the slope $\partial{H_{c2}}/\partial{T}$
-- are comparable in value to the those found in other
transition-metal-substituted \Ba\ series, indicating the robust
nature of  superconductivity induced by substitution of Group VIII
elements.
\end{abstract}

\maketitle

\section{Introduction}
The recent discovery of high-temperature superconductivity in
iron-based pnictide compounds has attracted much interest among the
condensed matter community, providing both a new angle for
understanding the physics of unconventional superconductivity in
other materials such as the copper-oxides, heavy-fermion
intermetallics, etc., and an entire new family of superconducting
materials of fundamental and technological interest. The highest
\tc\ achieved so far in these materials is $\sim 55$~K in
SmO$_{1-x}$F$_x$FeAs~\cite{Ren1} and
(Ba,Sr,Ca)FeAsF~\cite{Zhu,Cheng}. Oxygen-free FeAs-based compounds
with the ThCr$_2$Si$_2$-type (122) structure also exhibit
superconductivity with transition temperatures reaching $\sim 37$~K,
induced by chemical substitution of alkali or transition metal
ions~\cite{Sasmal,Rotter,Sefat,Leithe}, the application of large
pressures~\cite{Torikachvili,CaFe2As2,Alireza,Kumar}, or lattice
strain \cite{saha}.

It is widely believed that suppression of the magnetic/structural
phase transition in these materials, either by chemical doping or
high pressure, is playing a key role in stabilizing
superconductivity in the
ferropnicitides~\cite{bondangle,Kreyssig,Canfield}. For instance,
superconductivity has been induced by partial substitution of Fe by
other transition metal elements from the Fe, Co and Ni groups in
both the 1111~\cite{Sefat2,Wang1,Cao1} and 122
compounds~\cite{Sefat,Leithe,saha1}. For the 122 phase,
superconductivity has been induced by substituting Fe with not only
3$d$-transition metals such as Co and Ni, but also some of the 4$d$-
and 5$d$-transition metals.  Superconductivity with \tc\ as high as
25 K has been observed in \Bac~\cite{Ni,Chu} and \Barh~\cite{Ni3},
and near 20 K in \Ban~\cite{Li,Canfield} and \Bapd~\cite{Ni3}
compounds. Recently, Ru and Ir substitution for Fe were also shown
to induce superconductivity in polycrystalline \Sr\
samples~\cite{Schnelle,Han}, leaving only Os and Pt substitutions
from the Group VIII elements uninvestigated.

Here we report the first case of superconductivity induced by Pt substitution in the
FeAs-based family, presenting the observation of superconductivity at 23~K in \Bapt.
We present details of the synthesis and characterization of single crystals of
this material via single-crystal x-ray diffraction, electrical resistivity, magnetic susceptibility
and specific heat experiments.
%
\section{Experiment}
Single-crystalline samples of \Bap\ were grown using the FeAs
self-flux method~\cite{saha1}. Fe and Pt were first separately
pre-reacted with As via solid-state reaction of Fe (99.999\%)/Pt
(99.99\%) powder with As (99.99\%) powders in a quartz tube of
partial atmospheric pressure of Ar. The precursor materials were
mixed with elemental Ba (99.95\%) in the ratio of FeAs:PtAs:Ba =
$4-2x$:$2x$:1, placed in an alumina crucible and sealed in a quartz
tube under partial Ar pressure. The mixture was heated to
1150$^\circ$C, slow-cooled to a lower temperature and then quenched
to room temperature. Typical dimensions of as-grown single crystal
specimen of \Bapt\ are $0.1 \times 1 \times 2$~mm$^3$. Structural
properties were characterized by single-crystal X-ray-diffraction
and Rietfeld refinement (SHELXS-97). Chemical analysis was obtained
via both energy- and wavelength-dispersive X-ray spectroscopy,
showing proper stoichiometry in all specimens reported herein and no
indication of impurity phases. Resistivity ($\rho$) samples were
prepared using gold wire/silver paint contacts made at room
temperature, yielding typical contact resistances of $\sim$ 1
$\Omega$. Resistance measurements were performed using the standard
four-probe AC method, with excitation currents of 1~mA at higher
temperatures that were reduced to 0.3~mA at low temperatures to
avoid self-heating, all driven at 17 Hz in a Quantum Design PPMS
equipped with superconducting magnet. Magnetic susceptibility
($\chi$) and magnetization were measured using a Quantum Design
SQUID magnetometer, and specific heat was measured with a Quantum
Design PPMS cryostat using the thermal relaxation method.
%
%
%
\section{Results}
%
\subsection{Crystallographic Parameters}
Table~1 shows the crystallographic parameters determined by
single-crystal X-ray-diffraction at 250~K in \Bapt. A Bruker Smart
Apex2 diffractometer with MoK$\alpha$ radiation, a graphite
monochromator with monocarp collimator, and a CCD area detector were
used for this experiment. The structure was refined with SHELXL-97
software using 1033 measured reflections of which 115 were unique
and 108 observed. The final residuals were $R_1$=1.95\% for the
observed data and $wR_2$=4.46\% for all data. Fe and Pt atoms were
found to reside in the same site with a refined Fe:Pt ratio of
0.953(4):0.047(4), giving the exact formula
BaFe$_{1.906(8)}$Pt$_{0.094(8)}$As$_2$ from x-ray analysis.
Refinement data for \Ba\ determined by powder diffraction are
adopted from Ref.~\cite{rotter1} for comparison. As shown in
Table~1, the $c$-axis and the $c/a$ ratio shrink due to Pt
substitution, while the $a$-axis and the unit cell volume expands as
compared to undoped \Ba. Of note, the relative height of As above
the Fe lattice ($z$-parameter) and the corresponding As-Fe-As bond
angles change very little with Pt substitution.

\begin{table}
\caption{\label{tabl1}Crystallographic data for \Bapt\ determined by
single-crystal x-ray diffraction. The structure was solved and
refined using the SHELXS-97 software, yielding lattice constants
with residual factor $R$=1.95\%. Data for \Ba\ determined by powder
diffraction are adopted from Ref.~\cite{rotter1}.}
\footnotesize\rm
\begin{ruledtabular}
\begin{tabular}{lll}
&\Ba&\Bapt\\
\hline
Temperature&297~K&250~K\\
Structure&Tetragonal&Tetragonal\\
Space group&I4/mmm&I4/mmm\\
$a$($\mathrm{\AA}$)&3.9625(1)&3.9772(9)\\
$b$($\mathrm{\AA}$)&=$a$&=$a$\\
$c$($\mathrm{\AA}$)&13.0168(3)&12.988(6)\\
$V^3$($\mathrm{\AA}^3)$&204.38(1)&205.45(3)\\
$Z$ (formula unit/unit cell)&2&2\\
Density(g/cm$^3$)&---&6.673\\
Atomic parameters:&&\\
Ba&2$a$(0,0,0)&2$a$(0,0,0)\\
Fe/Pt&4d(1/2,0,1/4)&4$d$(1/2,0,1/4)\\
As&4e(0,0,$z$)&4$e$(0,0,$z$)\\
&$z$=0.3545(1)&$z$=0.35422(9)\\
Bond lengths ($\mathrm{\AA}$):&&\\
Ba-As($\mathrm{\AA}$)&3.382(1) x 8&3.3903(10) x 8\\
Fe-As($\mathrm{\AA}$)&2.403(1) x 4&2.4056(8) x 4\\
Fe-Fe/Pt($\mathrm{\AA}$)&2.802(1) x 4&2.8123(6) x 4\\
Bond angles (deg):&&\\
As-Fe-As&111.1(1) x 2&111.51(5) x 2\\
As-Fe-As&108.7(1) x 4&108.46(3) x 4\\

\end{tabular}
\end{ruledtabular}
\end{table}
%
%
\subsection{Electrical Resistivity}
%
%
\begin{figure}[tbh]
\begin{center}
\includegraphics[width=8.0cm]{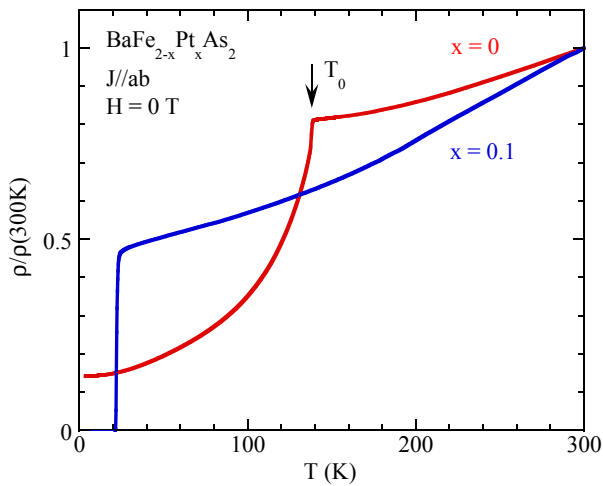}
\end{center}
\caption{\label{fig1}Temperature dependence of in-plane electrical
              resistivity of \Bap\ for $x$=0 and 0.1 normalized to
              300~K.  The arrow indicates the position
              of the antiferromagnetic transition associated with a structural transition
              at temperature $T_0$, defined by the kink in $x=0$
              data.}
\end{figure}
Figure~\ref{fig1}(a) presents the comparison of the in-plane
resistivity $\rho(T)$ of single crystals of \Bap\ with x=0 and 0.1
in zero applied magnetic field (data are presented after normalizing
to room temperature). As shown, $\rho(T)$ data for \Ba\ exhibit
metallic behavior, decreasing with temperature from 300~K before
exhibiting a sharp kink at $T_0=139$~K, where a structural phase
transition (from tetragonal to orthorhombic upon cooling) is known
to coincide with the onset of antiferromagnetic (AFM) order
\cite{rotter1}. For $x=0$, $\rho$ continues to decrease below $T_0$
without any trace of strain-induced superconductivity down to 1.8
K~\cite{saha}. The drop in $\rho(T)$ below $T_0$ has also been
observed in other 122 materials~\cite{Li,Leithe,Ni,Canfield}, and
likely arises due to the balance between the loss of inelastic
scattering due to the onset of magnetic order and the change in
carrier concentration associated with the transition at $T_0$. In
the $x=0.10$ sample, there is no indication of the $T_0$ transition
in $\rho(T)$ down to the superconducting transition that onsets at
\tc =23~K and drops to zero resistance by 21.5~K, indicating a
resistive transition width $\Delta T_c < 1.5$~K.
%
%
\begin{figure}[tbh]
\begin{center}
\includegraphics[width=8.0cm]{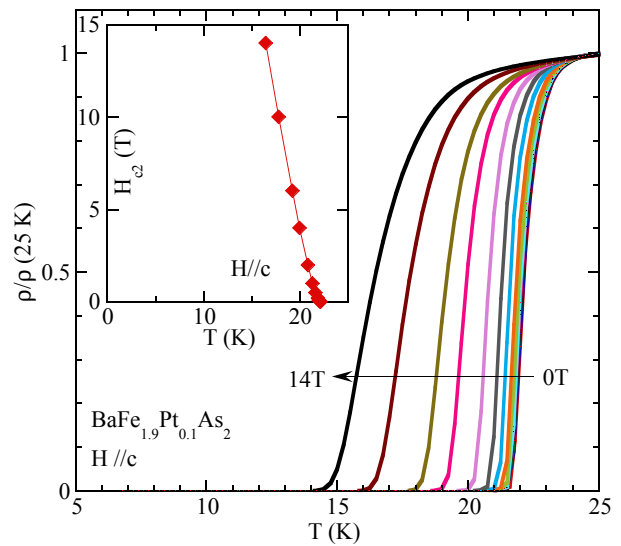}
\end{center}
  \caption{\label{fig2}In-plane electrical resistivity of \Bapt\ in magnetic
  fields (0, 0.01, 0.02, 0.03, 0.04,0.05, 0.1, 0.2, 0.5, 1, 2, 4, 6,10
and 14~T indicated by the direction of the arrow)  applied along the
$c$-axis, showing the suppression of the superconducting
     transition by increasing field. The resistivity is normalized to the normal state
     value (at 25~K) just above the transition for clarity. The inset shows upper critical field
     $H_{c2}$ of \Bapt, for $H \parallel c$ vs temperature. The points here denote the 50\% positions
     of the resistive transitions for each field from the main figure.}
\end{figure}
The suppression of the resistive superconducting transition of
\Bapt\ with the magnetic field $H$ applied parallel to the $c$-axis
is illustrated in Fig.~\ref{fig2}. The data are normalized to the
normal-state resistance above \tc\ for clarity. Applied magnetic
fields causes a tiny negative magnetoresistance at 25~K ([$\rho
(H=14~{\rm T})-\rho (H=0)$]/$\rho (H=0)\sim -0.35\%$). The
superconducting upper critical field $H_{c2}(T)$, as determined by
the 50\% resistive transition temperature for each field, is shown
in the inset of Fig.~\ref{fig2}. The slope
$\partial{H_{c2}}/\partial{T}$ is -2.8~T/K in the range $T<20$~K,
which is comparable to values reported for other transition
metal-doped FeAs-based superconductors~\cite{butch1}. This agreement
is rather remarkable, given that these superconductors have values
of \tc\ ranging from 10~K to more than 30~K~\cite{butch1}. A simple
estimate using the Werthamer-Helfland-Hohenberg (WHH) approximation
$H_{c2}(0) \simeq 0.691\frac{\partial{H_{c2}}}{\partial{T}}T_c$
yields a value of $\sim 45$~T for the orbital $H_{c2}(0)$. However,
previous studies (See Ref.~\cite{butch1}) have shown a more linear
dependence of $H_{c2}(T)$ toward lower temperature, which would
extrapolate to a value $H_{c2}(0)\approx 65$~T for \Bapt. In any
case, the response of the superconducting state to applied $H$ seems
insensitive to whether superconductivity has been stabilized by
different transition metals substitution, applied pressure, or
presumably strain, in the case of the undoped parent compounds. In
contrast, hole-doped \Sr\ and \Ba\ feature larger values of
$\partial{H_{c2}}/\partial{T}$~\cite{butch1}.
%
%
\subsection{Magnetic Susceptibility and Magnetization}
%
\begin{figure}[tbh]
\begin{center}
\includegraphics[width=8.0cm]{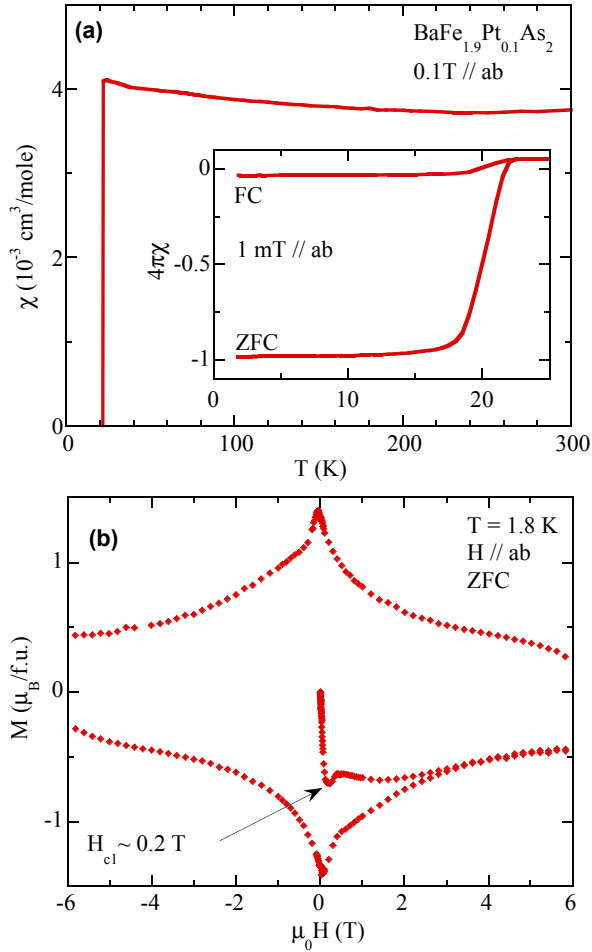}
\end{center}
  \caption{\label{fig3}(a) Temperature dependence of the magnetic susceptibility $\chi(T)$ in
  the \Bapt\ measured with 100~mT field applied parallel to the
  crystallographic basal plane following zero-field-cooled (ZFC) conditions.
  Inset: volume magnetic susceptibility at 1~mT for both ZFC and field-cooled (FC) conditions.
  (b) Magnetization of \Bapt\ as a function of applied field at 1.8 K, with H applied in-plane.
  The loop has a large open-ended area indicating values of $H_{c2}$ well exceeding 6~T.
  Demagnetization effects due to sample geometry have been corrected for both
  $\chi (T)$ and $M(H)$ data.}
\end{figure}
Figure~\ref{fig3}(a) presents the temperature dependence of magnetic
susceptibility $\chi$ of \Bapt\ measured under zero-field-cooled
(ZFC) conditions by applying a magnetic field of 100~mT along the
$ab$-plane at low temperatures. As shown, $\chi$ is nearly
temperature-independent down to 23~K with no indication of a
magnetic/structural transition. Below 23~K, $\chi (T)$ sharply drops
to negative values due to Meissner screening. The inset of
Fig.~\ref{fig3} presents volume susceptibility 4$\pi \chi$ for both
at ZFC and field-cooled (FC) conditions under a 1.0 mT magnetic
field applied along the $ab$-plane at low temperatures in order to
compare the level of diamagnetic screening due to superconductivity.
There is a relatively sharp drop of ZFC susceptibility from positive
to negative value below \tc=23~K (onset) consistent with the
resistivity data. As shown, the superconducting volume fraction, as
estimated by the fraction of full diamagnetic screening ($4\pi
\chi=-1$), reaches 100\% at $\sim 17$~K, indicating full Meissner
effect.

In order to study the magnetic response of the superconducting
state, the magnetization was measured as a function of field.
Fig.~\ref{fig3}(b) shows the isothermal ($T$=1.8K) magnetization
$M(H)$ measured from ZFC conditions. Magnetization is nonlinear and
irreversible with $H$ due to superconductivity, and it is evident
that the apparent value of $H_{c2}$ well exceeds 6~T, consistent
with our resistivity studies. A minimum is observed at low $H$
region in the virgin curve as identified by the diagonal arrow in
Fig.~\ref{fig3}(b), indicating the lower limit of  a superconducting
lower critical field value $H_{c1}=200$~mT.
%
\begin{figure}[tbh]
\begin{center}
\includegraphics[width=8.0cm]{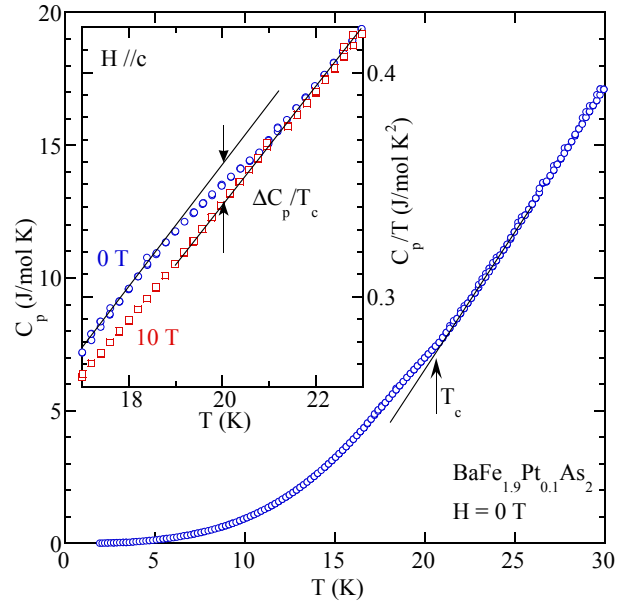}
\end{center}
  \caption{\label{fig4}Temperature-dependent heat capacity data for
\Bapt, showing a distinct feature centered at $\sim 20$~K, indicated
by drawing a linear line through the data points and an upward
arrow, that is consistent with the onset of superconductivity at
$T_c=23$~K determined from resistivity and magnetic susceptibility
measurements. Inset: zoom of temperature dependence of $C_p/T$ near
the superconducting transition. Lines are fits to the data above and
below \tc, with the arrows indicating the position where $\Delta
C_p/T_c$ is determined. Data for $H=10$~T are also plotted for
comparison.}
\end{figure}
%
%
\subsection{Specific Heat}
Specific heat measurements were performed to verify the bulk
thermodynamic nature of the superconducting transition in \Bapt.
Temperature dependent heat capacity data in zero magnetic field for
\Bapt\ are plotted in fig.~\ref{fig4}. An abrupt shift in the smooth
$C_p(T)$ curve below 21~K, indicated by the arrow, is consistent
with the superconducting phase transition observed in both
resistivity and low-field magnetization data. Both the
superconducting temperature and the upper critical fields in \Bapt\
are high, thus making a reliable estimate of the normal state
electronic specific heat $\gamma$ difficult. For this reason we have
determined $\Delta C_p/T_c$ rather than the more traditional
quantity $\Delta C_p/\gamma T_c$. Due to finite widths of
superconducting transitions $\Delta C_p/T_c$ and $T_c$ values are
determined from plots of $C_p/T$ vs $T$ shown in the inset of
fig.~\ref{fig4} in an enlarged view using an isoentropic
construction as done previously~\cite{budko1}. Data for $H=$10~T
(red square symbols), in which the anomaly due to superconductivity
has been shifted to lower temperature, serve as the lower line for
this analysis. The vertical distance between up and down arrow gives
the value of $\Delta C_p/T_c \simeq$ 20 mJ/mol~K. Assuming the BCS
weak-coupling approximation $\Delta C_p/\gamma T_c$=1.43 and 100\%
superconducting volume, the value of $\gamma$ can be estimated to be
about 14 mJ/mol K$^2$, which is comparable to that found in other
transition metal-doped \Ba\ superconductors~\cite{Ni2}. For $x=$0 in
Ref.~\cite{rotter1}, a $C_p /T$ vs $T^2$ plot between 3.1~K and 14~K
gives $\gamma = $16(2)~mJ/mol~K$^2$ corresponding to a Debye
temperature of $\Theta_D =$ 134(1)~K. In Ref.~\cite{budko1}, the
values of $\Delta C_p/T_c$ measured for K, Co, Ni, Rh, and Pd-doped
\Ba\ superconductors have been shown to scale with $T_c$, regardless
of the value of \tc\ or the relative doping position (under- or
over-doped) with respect to maximum \tc. Surprisingly, the
corresponding value of $\Delta C_p/T_c \simeq$ 20 mJ/mol~K taken at
20~K for \Bapt\ also falls in line with this trend, expanding this
interesting relation to include another $5d$-transition metal-doped
system.
%
%
\section{Discussion}
Although the detailed phase diagram is yet unknown, the
superconducting properties of this new member of the superconducting
FeAs-based materials look to be strikingly similar to those observed
in the other related compounds. The widely perceived picture is that
pairing occurs through the interpocket scattering of electrons via
exchange of antiferromagnetic spin fluctuations~\cite{Mazin}. By
doping electrons or holes into the parent phase, magnetic order is
gradually destroyed and the short-range order provides a wide
spectrum of spin fluctuations which may be responsible for pairing
between electrons. Alternatively, magnetic order and
superconductivity may compete to gap similar parts of the Fermi
surface, with superconductivity only appearing when magnetic order
is suppressed. It is certain that superconductivity is associated,
directly or indirectly, with the suppression of magnetic order in
the FeAs-based 122 systems, and an understanding of the generalized
phase diagram must be an integral part of an explanation of the
physics of these materials.

This picture can certainly give a qualitative explanation for the
generally similar occurrence of superconductivity via doping of
different species of Group VIII elements (including Co, Rh, Ir, Ni,
Pd, and now Pt) into the 122 parent compounds. However, subtle
details and differences among these different series may hold
important information regarding the specific mechanism(s) by which
magnetic order is suppressed and superconductivity is optimized. For
instance, in \Bac\ the maximum \tc\ is found at $x\simeq 0.15$
~\cite{Chu,Wang2}, whereas in \Ban\ the maximum \tc\ occurs at
approximately $x\simeq 0.10$ ~\cite{Li,Canfield}. This appears to be
consistent with simple $d$-electron counting, and hence charge
doping, however it is not clear that simple scaling of different
phase diagrams by electron count works in all cases \cite{saha1}.
Another interesting aspect of the superconductivity in 122 materials
is the similarity of maximum $T_c$ values, typically reaching
20-25~K regardless of the transition metal
substituent~\cite{Ni2,Han}. While this is thus far true in all
Ba-based compounds, the trend is broken in Sr-based 122 systems such
as SrFe$_{2-x}$Pd$_{x}$As$_{2}$~\cite{Han} and
SrFe$_{2-x}$Ni$_{x}$As$_{2}$~\cite{saha1}, which both exhibit
reduced maximum values of \tc\ closer to $\sim 10$~K. While the
value of \tc\ in \Bapt\ is in line with most other transition-metal
doped \Ba\ superconductors, it is slightly higher than the maximum
\tc\ of $\sim 18$~K found in the closely related series \Bapd\
\cite{Ni3}. Although differences in maximal \tc\ values could arise
for many different reasons, intrinsic variations of \tc\ in
different doping series may be an important indicator of the nature
of pairing in this family of materials. Because it is known that
annealing treatments of as-grown crystals of \Srn\ result in a
significant enhancement in superconducting transition
temperatures~\cite{saha2}, it will be important to investigate the
role of crystal quality on variations in \tc\ values in these
systems.

Finally, because Os is now the only remaining element of the Fe, Co
and Ni transition metal groups to be investigated, it will be
surprising if Os substitution does not also induce superconductivity
in \Ba\ or \Sr\ systems. Future work will investigate this question.
\section{Summary}
In summary, single crystals of the 5$d$-transition metal
Pt-substituted \Ba\ were successfully synthesized and shown to
become bulk superconductors, leaving only one remaining element in
the Group VIII transition metals to be shown to induce
superconductivity in the iron-pnictide family of materials.
Transport, magnetic and thermodynamic studies have revealed
superconductivity below $T_c=23$~K in \Bapt\ to be of a bulk nature
and robust against applied magnetic field, with estimates of upper
critical field values near $\sim 65$~T. The similarity in
superconducting properties between the substitution series of
different transition metals suggests that similar underlying physics
is at play in stabilizing superconductivity in this family of
materials.
%
%
\vskip 0.5cm
\begin{center}
ACKNOWLEDGEMENTS
\end{center}
The authors acknowledge B.~W.~Eichhorn for experimental assistance,
and N.P.B. acknowledges support from a CNAM Glover fellowship. This
work was supported by AFOSR-MURI Grant FA9550-09-1-0603.




\begin{thebibliography}{11}

\bibitem{Ren1}Ren Z -A, Lu W, Yang J, Yi W, Shen X -L, Zheng-Cai, Che G -C, Dong X -L,
Sun L -L, Zhou F, and Zhao Z -X, 2008 {\it Chin. Phys. Lett.} {\bf
25} 2215.

\bibitem{Zhu}Zhu X, Han F, Cheng P, Mu G, Shen B, and Wen H H, 2009 {\it Europhys.
Lett.} {\bf 85} 17011.

\bibitem{Cheng}Cheng P, Shen B, Mu G, Zhu X, Han F,
Zeng B, and Wen H H, 2009 {\it Euro Phys. Lett.} {\bf 85} 67003.


\bibitem{Sasmal}Sasmal K, Lv B, Lorenz B, Guloy A M, Chen F, Xue Y -Y, and Chu C -W, 2008 {\it Phys. Rev. Lett.} {\bf 101} 107007.

\bibitem{Rotter}Rotter M, Tegel M, and Johrendt D, 2008 {\it Phys. Rev. Lett.} {\bf 101} 107006.

\bibitem{Sefat}Sefat A S, Jin R, McGuire M A, Sales B C, Singh D J, and Mandrus D, 2008 {\it Phys. Rev. Lett.} {\bf 101} 117004.

\bibitem{Leithe}Leithe-Jasper A, Schnelle W, Geibel C, and Rosner H, 2008 {\it Phys. Rev. Lett.} {\bf 101} 207004.

\bibitem{Torikachvili}Torikachvili M S, Bud'ko S L, Ni N, and Canfield P C, 2008 {\it Phys. Rev. Lett.} {\bf 101} 057006.

\bibitem{CaFe2As2}Park T, Park E, Lee H, Klimczuk T, Bauer E D, Ronning F, and  Thompson J D, 2008 {\it J. Phys.: Condens. Matt.} {\bf 20} 322204.

\bibitem{Alireza}Alireza P L, Chris-Ko Y T, Gillett J, Petrone C M, Cole J M,
Lonzarich G G, and Sebastian S E, 2009 {\it J. Phys.: Condens.
Matt.} {\bf 21}, 012208.

\bibitem{Kumar}Kumar M, Nicklas M, Jesche A, Caroca-Canales N, Schmitt M, Hanfland M,
Kasinathan D, Schwarz U, Rosner H, and Geibel C, 2008 {\it Phys.
Rev. B} {\bf 78} 184516.

\bibitem{saha} Saha S R, Butch N P,  Kirshenbaum K, and  Paglione Johnpierre, 2009 {\it Phys. Rev. Lett.}, {\bf 103} 037005.

\bibitem{bondangle}Lee C -H, Iyo A, Eisaki H, Kito H, Fernandez-Diaz M T, Ito T, Kihou K,
Matsuhata H, Braden M, and Yamada K, 2008 {\it J. Phys. Soc. Jpn.}
{\bf 77} 083704.

\bibitem{Kreyssig}Kreyssig A, Green M A, Lee Y, Samolyuk G D, Zajdel P,
Lynn J W, Bud'ko S L, Torikachvili M S, Ni N, Nandi S, J B Leao,
Poulton S J, Argyriou D N, Harmon B N, McQueeney R J, Canfield P C,
and Goldman A I, 2008 {\it Phys. Rev. B} {\bf 78} 184517.

\bibitem{Canfield}Canfield P C, Bud'ko S L, Ni N, Yan J Q,
Kracher A, 2009 {\it Phys. Rev. B} {\bf 80} 060501(R).

\bibitem{Sefat2}Sefat A S, Huq A, McGuire M A, Jin R,
Sales B C, Mandrus D, Cranswick L M D, Stephens P W, and Stone K H,
2008 {\it Phys. Rev. B} {\bf 78} 104505.

\bibitem{Wang1}Wang C, Li Y K, Zhu Z W, Jiang S, Lin X, Luo Y K, Chi S,
 Li L J, Ren Z, He M, Chen H, Wang Y T, Tao Q, Cao G H, and Xu Z A, 2009 {\it Phys. Rev. B} {\bf 79} 054521.

\bibitem{Cao1}Cao G, Jiang S, Lin X, Wang C, Li Y, Ren Z, Tao Q, Feng C, Dai J,  Xu Z A, Zhang F -C, 2009 {\it Phys. Rev. B} {\bf 79} 174505.

\bibitem{saha1}Saha S R, Butch N P,  Kirshenbaum K, and  Paglione Johnpierre, 2009 {\it Phys. Rev. B} {\bf 79} 224519.

\bibitem{Ni}Ni N, Tillman M E, Yan J -Q, Kracher A, Hannahs S T, Bud'ko S L, and
Canfield P C, 2008 {\it Phys. Rev. B} {\bf 78}, 214515.

\bibitem{Chu}Chu J -H, Analytis J G, Kucharczyk C, and Fisher I R, 2009 {\it Phys. Rev. B} {\bf 79} 014506.

\bibitem{Ni3}Ni N, Thaler A, Kracher A, Yan J -Q,  Bud'ko S L, and Canfield P C, 2009 {\it Phys. Rev. B} {\bf 80} 024511.

\bibitem{Li}Li L J, Wang Q B, Luo Y K, Chen H, Tao Q, Li Y K, Lin X, He M, Zhu Z W,
Cao G H, and Xu Z A, 2009 {\it New J. Phys.} {\bf 11} 025008.

\bibitem{Schnelle}Schnelle W, Leithe-Jasper A, Gumeniuk R, Burkhardt U, Kasinathan D,
Rosner H, 2009 {\it Phys. Rev. B} {\bf 79} 214516.

\bibitem{Han}Han F, Zhu X, Cheng P, Mu G, Jia Y, Fang L, Wang Y, Luo H, Zheng B, Shan L, Ren C, Wen H H,
2009 {\it Phys. Rev. B.} {\bf 80} 024506.

\bibitem{rotter1} Rotter M, Tegel M, Johrendt D, Schellenberg I, Hermes W, and P\"{o}ttgen R, 2008  {\it Phys. Rev. B} {\bf 78} 020503(R).

\bibitem{butch1} Butch N P, Saha S R, Zhang X, Kirshenbaum K, Greene R and  Paglione Johnpierre, 2010 {\it Phys. Rev. B} {\bf 81} 024518.

\bibitem{budko1}Bud'ko S L , Ni N, and Canfield P C, 2009 {\it Phys. Rev. B} {\bf 79} 220516(R).

\bibitem{Ni2}Ni N, Thaler A, Yan J-Q, Kracher A, Bud'ko S L, and Canfield P C, 2009 {\it Phys. Rev. B} {\bf 80} 214511.

\bibitem{Mazin}Mazin I I, Schmalian J, 2009 {\it Physica C} {\bf 469} 614.

\bibitem{Wang2}Wang X F, Wu T, Wu G, Liu R H, Chen H, Xie Y L and Chen X H, 2009
{\it New J. Phys.} {\bf 11} 045003.

\bibitem{saha2} Saha S R, Butch N P,  Kirshenbaum K, and  Paglione J, 2010 {\it Physica C} at press
(arXiv:0908.4095).



\end{thebibliography}
\end{document}